\documentclass[aps,twocolumn,amsmath,amssymb,showpacs,prb,superscriptaddress,unsortedaddress]{revtex4}

\usepackage{epsf}
\usepackage{graphicx}

\begin{document}

\title{Anomalous asymmetry of the Fermi surface in the YBa$_2$Cu$_4$O$_{8}$ high temperature superconductor revealed by Angle Resolved Photoemission Spectroscopy}

\author{Takeshi Kondo}
\affiliation{Ames Laboratory and Department of Physics and Astronomy, Iowa State University, Ames, IA 50011, USA}

\author{R.~Khasanov}
 \affiliation{Laboratory for Muon Spin Spectroscopy, Paul Scherrer
Institut, CH-5232 Villigen PSI, Switzerland}

\author{Y.~Sassa}
\affiliation{Laboratory for Neutron Scattering, ETH Z\"urich and Paul Scherrer Institute, CH-5232 Villigen PSI, Switzerland}

\author{A.~Bendounan}
\affiliation{Swiss Light Source, Paul Scherrer Institute, CH-5232 Villigen PSI, Switzerland}

\author{S.~Pailhes}
\affiliation{Laboratoire L\'eon Bril louin, CEA-CNRS, CEA-Saclay, 91191 Gif-sur-Yvette, France}

\author{J.~Chang}
\affiliation{Laboratory for Neutron Scattering, ETH Z\"urich and Paul Scherrer Institute, CH-5232 Villigen PSI, Switzerland}

\author{J.~Mesot}
\affiliation{Laboratory for Neutron Scattering, ETH Z\"urich and Paul Scherrer Institute, CH-5232 Villigen PSI, Switzerland}

\author{H.~Keller}
\affiliation{Physik-Institut der Universit\"{a}t Z\"{u}rich,
Winterthurerstrasse 190, CH-8057 Z\"urich, Switzerland}

\author{N.~D.~Zhigadlo}
\affiliation{Laboratory for Solid State Physics ETH Z\"urich, CH-8093 Z\"urich, Switzerland}

\author{M.~Shi}
\affiliation{Swiss Light Source, Paul Scherrer Institute, CH-5232 Villigen PSI, Switzerland}

\author{Z.~Bukowski}
\affiliation{Laboratory for Solid State Physics ETH Z\"urich, CH-8093 Z\"urich, Switzerland}

\author{J.~Karpinski}
\affiliation{Laboratory for Solid State Physics ETH Z\"urich, CH-8093 Z\"urich, Switzerland}

\author{A.~Kaminski}
\affiliation{Ames Laboratory and Department of Physics and Astronomy, Iowa State University, Ames, IA 50011, USA}

\date{\today}
\begin{abstract}
We use microprobe Angle-Resolved Photoemission Spectroscopy ($\mu$ARPES) to study the Fermi surface and band dispersion of the CuO$_{2}$ planes in the high temperature superconductor, YBa$_2$Cu$_4$O$_{8}$. We find a strong in-plane asymmetry of the electronic structure between directions along $a$ and $b$ axes. The saddle point of the antibonding band lies at a significantly higher energy in the $a$-direction ($\pi$,0) than the $b$-direction (0,$\pi$), whereas the bonding band displays the opposite behavior. We demonstrate that the abnormal band shape is due to a strong asymmetry of the bilayer band splitting, likely caused by a non-trivial hybridization between the planes and chains. This asymmetry has an important implication for interpreting key properties of the Y-Ba-Cu-O (YBCO) family, especially the superconducting gap, transport and results of inelastic neutron scattering.
\end{abstract}

\pacs{74.25.Jb, 74.72.Hs, 79.60.Bm}

\maketitle

It is commonly accepted that the conduction electrons/holes in the CuO$_2$ planes play an essential role in determining the electronic properties of the high-$T_c$ superconductors. It is however not totally understood how the charge carriers arrange themselves and interact with each other. Strong $a$-$b$ axis asymmetry of both the normal- and superconducting-state electronic properties have been observed in the Y-Ba-Cu-O (YBCO) family especially in YBa$_2$Cu$_3$O$_{7-\delta}$ (Y123) and YBa$_2$Cu$_4$O$_{8}$ (Y124), which possess double CuO$_{2}$ planes sandwiched between single and double CuO chains, respectively.  The London penetration depth ($\lambda$) measured by muon-spin rotation shows a strong asymmetry ($\lambda _a /\lambda _b  \approx 1.2$ in Y123 \cite{Ager,Rustem_Y123} and $\lambda _a /\lambda _b  \approx 1.5$ in Y124 \cite{Rustem_Y124}). Here the $a$- and $b$-axis directions are perpendicular to and along the chains, respectively. Angle Resolved Photoemission Spectroscopy \cite{Lu} observes a 50$\%$ larger gap along the chain direction in Y123. Angle-Resolved Electron tunneling in Y123/Au/Nb junctions also indicates a superconducting gap ($\Delta$) asymmetry of $\Delta _b /\Delta _a  \approx 1.5$ \cite{Smilde}.  Most recent reports using a similar technique, however, show only 20$\%$ difference \cite{Kirtley}. Many theoretical models have been proposed in order to explain these results. These include the formation of striped phases \cite{Kivelson}, orthorhombicity of the crystal structure \cite{Zhou}, proximity coupling between chains and planes \cite{Morr}, broken time-reversal symmetry \cite{Varma}, and  an admixture of $d_{x^2-y^2}$+s pair states \cite{Tsuei}.
The spin dynamics in Y123, observed in inelastic neutron scattering (INS) experiments, also shows a significant $a$-$b$ axis asymmetry in the vicinity of the wavevector (1/2,1/2) \cite{Mook,Arai,Bourges,Hinkov}. Mook $et.\ al.$ \cite{Mook2} report almost one-dimensional patterns and suggest that this is evidence for existing stripe phases \cite{Kivelson}.  In contrast, recent INS results on Y123 show that the spin fluctuations are two-dimensional, although they have a strong $a$-$b$ asymmetry \cite{Hinkov}. In the latter case, a Fermi-liquid-based scenario might be more relevant for explaining the asymmetric spin dynamics \cite{Zhou,Eremin,Schnyder}.

Part of the controversy arises due to a   lack of detailed band structure measurements in YBCO, since most  ARPES experiments performed on this material have focused on YBa$_2$Cu$_3$O$_{7-\delta}$ (Y123) variety, which has fractional oxygen stoichiometry in the chains. These samples do not cleave well and suffer complications due to the ordering of oxygen in the chains. The resulting unstable sample surfaces makes ARPES spectra difficult to interpret \cite{Lu}. Although band calculations predict a significant $a-b$ axis asymmetry of the band structure in CuO$_2$ planes due to a plane-chain coupling, that is often ignored when interpreting the electronic properties because of a lack of the experimental evidence supporting these predictions. Until now, there have been no observation of $a-b$ axis asymmetry in the Fermi surface of the CuO$_2$ planes by angle-resolved photoemission spectroscopy (ARPES) in YBCO \cite{Campuzano,Gofron,Shen1,Lu,Borisenco_YBCO,Nakayama,Kondo}.  We chose YBa$_2$Cu$_4$O$_{8}$ (Y124) for studying the band structure by ARPES, because Y124 has a fixed oxygen stoichiometry that results in a much more stable surface than that of Y123 and avoids problems due to oxygen ordering in chains. We have previously demonstrated that the plane and chain bands can be distinguished by using both a small UV beam (50-100$\mu$m) and a tunable incident photon energy \cite{Kondo}. This lead to observation of significant differences in the momentum dependence of the bilayers splitting between Bi2212 and YBCO. Since it was sufficient for this purpose to acquire data only along nodal and $a$-axis directions, we could not make statements about the $a-b$ axis anisotropy. In this letter, we compare the band structure along $a$ and $b$ axis with much improved data quality due to use of high flux beamline and longer acquisition times. We found a small but a significant $a-b$ axis asymmetry in the band structure, which is mainly due to the variation of the bilayer band splitting being about 40$\%$ larger along the $a$-direction. Our results suggest that the coupling between the planes and chains and the consequent asymmetry of the band structure represents a crucial ingredient in our understanding of the origin of the asymmetries in the electronic and magnetic properties of YBCO systems.

\begin{figure}
\includegraphics[width=3.5 in]{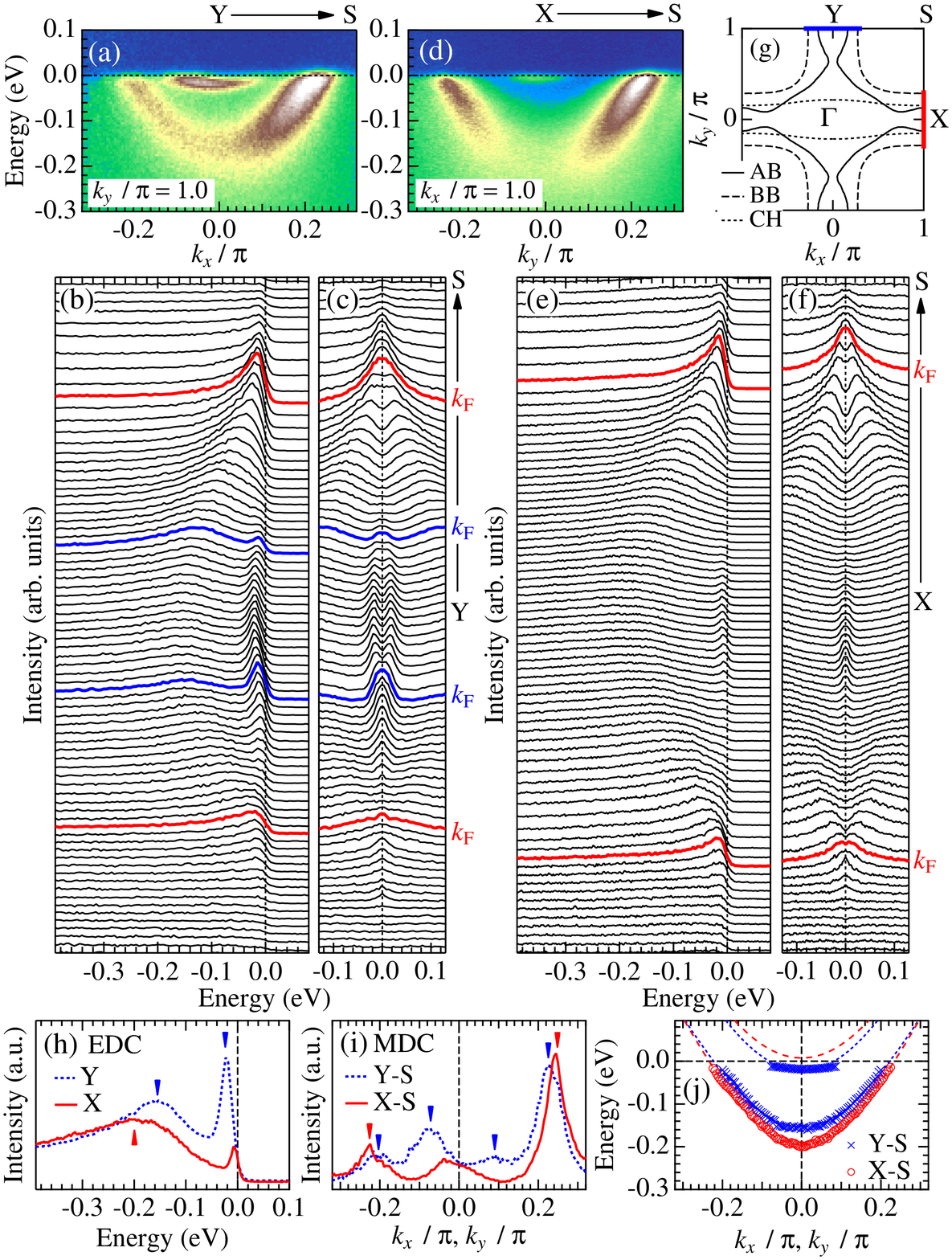}
\caption{(Color online) ARPES intensity of a plane domain and corresponding EDCs along Y-S (a,b) and X-S (d,e) indicated in (g). (c,f) Symmetrized EDCs of (b) and (e) close to the Fermi level.
(g) Fermi surfaces of antibonding-, bonding-, and chan-band (AB, BB, and CH, respectively) obtained by band calculation \cite{Anderson}. (h) EDCs at X and Y. (i) MDCs at the Fermi level along Y-S and X-S. (j) Band dispersions along Y-S and X-S determined from EDC peak positions of (b) and (e). Dotted blue lines and dashed red lines represent tight binding fits \cite{TB} along Y-S and X-S, respectively.}
\label{fig1}
\end{figure}

\begin{figure}
\includegraphics[width=3.5 in]{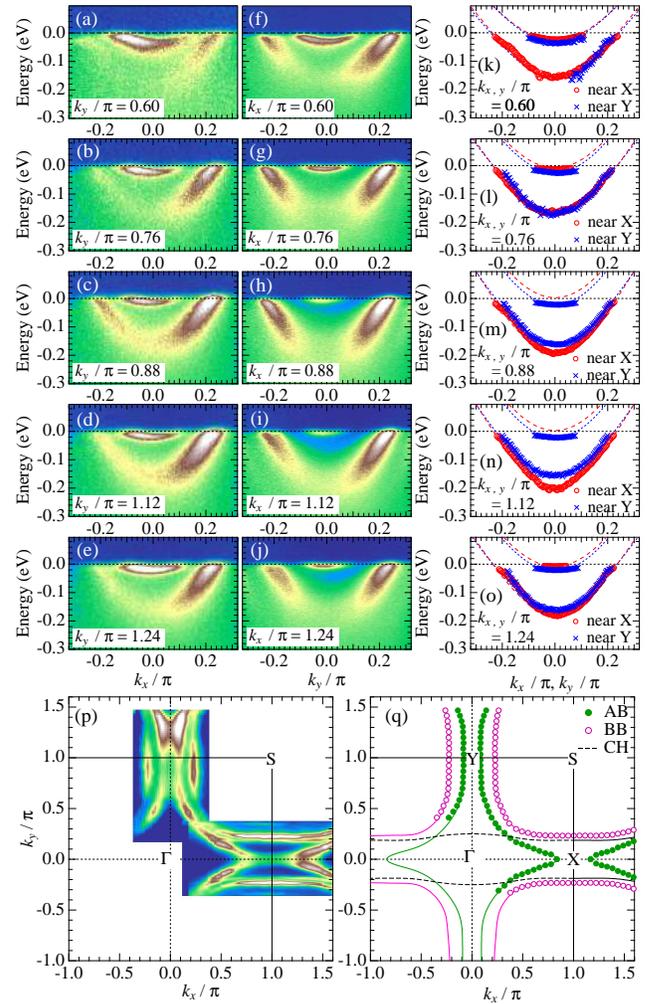}
\caption{(Color online) ARPES intensity at momentum cuts parallel to Y-S (a-e) and X-S (f-j). (k-o) Band dispersion determined from the EDC peak positions and tight binding fits \cite{TB}. (p) ARPES intensity integrated within $\pm10$meV centered at the Fermi level as a function of $k_x$ and $k_y$.  (q) Fermi surface plots of antibonding-band (AB, green filled circle) and bonding-band (BB, purple open circle) determined from the peak position of the MDCs at the Fermi level.}
\label{fig2}
\end{figure}

Untwinned high quality single crystals of underdoped YBa$_{2}$Cu$_{4}$O$_{8}$ ($T_c\simeq80$K) with a sharp superconducting transition ($\Delta T_c\simeq3$K) were grown by the self-flux method under high oxygen pressure \cite{Karpinski}. The ARPES experiments were carried out using a Scienta SES2002 hemispherical analyzer mounted on the SIS beamline of the Swiss Light Source (SLS). As a reference for the Fermi energy, we used the spectral edge position
of evaporated Au in electrical contact with the sample. The energy and angular resolutions were 20 meV and $0.1^\circ$, respectively. All spectra were measured at 25K using a 33eV photon energy. In order to observe both $a$- and $b$-axis directions with an identical ARPES setting (photon polarization and geometry of analyzer and sample), we rotated the samples by 90$^\circ$ following the same micro-domain on the sample surface. We measured several samples and always obtained similar results.

Figures 1 (a,b) and (d,e) show ARPES data (intensity maps and corresponding energy distribution curves (EDCs)) measured along Y-S and X-S, respectively. Those momentum cuts are illustrated in panel (g) along with the Fermi surface of the planes and chains obtained by band calculation \cite{Anderson}. Around Y (chain direction), both dispersions of the antibonding (AB) and bonding (BB) bands, attributed to the CuO$_2$ bilayer splitting, are clearly observed at higher and lower energies, respectively.
Note that, due to matrix element effects, the ARPES intensity at positive $k_x$ is stronger than at negative $k_x$ in the lower energy band. This $k_x$ dependence becomes opposite for the higher energy band. Such anti-correlation is typical for bonding and antibonding bands due to the orthogonality of their wavefunctions \cite{Bansil}. In contrast to the Y direction, the AB signal is very weak close to X (the direction perpendicular to the chains), whereas the BB is clearly seen.
The difference between the two antinodal cuts is more clearly illustrated by plotting the EDCs taken at X and Y in Fig.1 (h), and the momentum distribution curves (MDCs) at the Fermi level in Fig.1 (i).
We find that the BB dispersion around X displays a deeper bottom (by $\sim$40meV) and a wider Fermi closing than that around Y. Although a large peak corresponding to an energy state of AB is seen below the Fermi level (around $-20$meV) in the EDC at Y (panel (h)), the peak edge of a weak AB signal appears to be pinned to the $E_F$ at X.
We used the symmetrization method \cite{Norman} as well as the MDC peak position to determine the Fermi crossing points; EDCs are reflected about the Fermi level and added to the unreflected ones. This technique removes the Fermi function and enables us to immediately identify the Fermi crossing.
The symmetrized EDCs of Fig. 1 (b) and (e) are plotted in the Fig. 1 (c) and (f), respectively. When the dispersion crosses $E_F$, two peaks in the symmetrized EDC due to presence of low-lying energy states merge into a single one at Fermi momentum.
This is clearly seen both for AB and BB with moving away from Y along Y-S (panel (c)). 
The superconducting gap is not observed in these spectra because the YBCO sample surface is known to be overdoped after cleaving \cite{Kondo,DamascelliNP}. 
Along X-S (panel (f)), the BB clearly shows a Fermi crossing. On the other hand, the symmetrized EDC very close to X has only one peak, which indicates that there is no Fermi crossing in AB along this cut. The shape of symmetrized EDC is sensitive to the value of experimental energy resolution \cite{Mesot}. However we can confirm this from a behavior of the spectral intensities along $\Gamma$-X shown in Fig. 3(d); the intensity sharply decreases toward X, which indicates that the AB goes beyond  $E_F$ near X
and  only a small spectral tail is seen on the occupied side. In Fig. 1 (j),  we summarize the AB and BB dispersions of the occupied states determined from the peak position of the EDCs in panels (b) and (e) along with tight binding fitting curves \cite{TB}. Although our data represents the overdoped state, the band anisotropy reported here is expected to be valid for other doping levels because it is well known that doping of cuprates amounts to a rigid band shift \cite{Damascelli}.

\begin{figure}
\includegraphics[width=3.5 in]{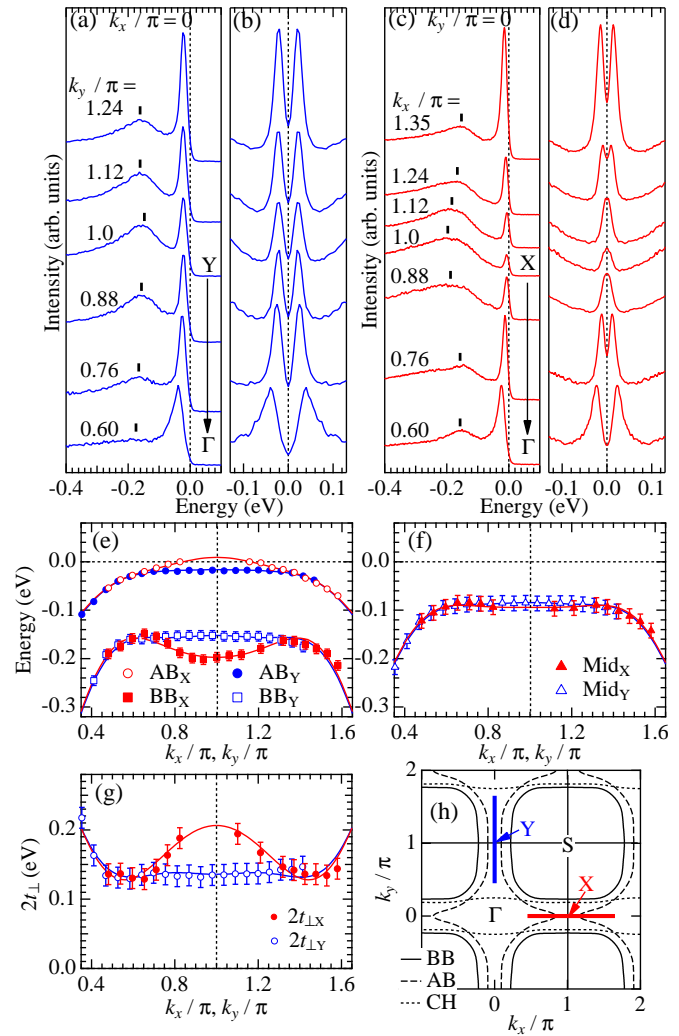}
\caption{(Color online) EDCs and the symmetrized ones measured at momenta along $\Gamma$-Y (a,b) and $\Gamma$-X (c,d) indicated in (h). (e) Dispersion of antibonding- and bonding-band along $\Gamma$-X (AB$_{\rm{X}}$ and BB$_{\rm{X}}$) and $\Gamma$-Y (AB$_{\rm{Y}}$ and BB$_{\rm{Y}}$) determined from peak positions of EDCs and symmetrized EDCs. Solid lines represent tight binding fits \cite{TB}. (f) The midpoints of AB and BB along $\Gamma$-X  (Mid$_{\rm{X}}$) and $\Gamma$-Y (Mid$_{\rm{Y}}$). (g) Energy of bilayer band splitting along $\Gamma$-X ($2t_\bot$$_{\rm{X}}$) and $\Gamma$-Y  ($2t_\bot$$_{\rm{Y}}$). (h) Brillouin zone.}
\label{fig3}
\end{figure}

Figure 2 (a-e) and (f-j) show ARPES intensity maps around Y and X measured along various momentum cuts parallel to Y-S and X-S, respectively.
Around X (panel (f-j)), the ARPES intensity near $E_F$ rapidly increases while moving away from the antinode due to the appearance of AB on the occupied side. This is in contrast to the strong intensity of the AB observed for all momentum cuts around Y (panels (a-e)).
We determine both AB and BB dispersions from the EDC peak positions, and plot the results near Y and X in Fig. 2(k-o); each panel is for two momentum cuts at equal distance from X and Y. The difference in the dispersions between the two regions reaches a maximum close to the antinodes, then decreases toward the nodes. The strong $a$-$b$ axis asymmetry of the Fermi surfaces is further visualized in Fig. 2(p) by plotting the ARPES intensity integrated within $\pm 10$meV of $E_F$ as a function of $k_x$ and $k_y$. In Fig. 2(q), we plot the Fermi crossing points extracted from the MDC peak positions at $E_F$. We find an interesting topology of the Fermi surface which appears to be hole-like (centered around S) close to Y and electron-like (centered around $\Gamma$) close to X. 
We should note that the previously reported data \cite{Kondo} was consistent with hole-like Fermi surface close to X. This difference is most likely due to small variation of doping for different batches of samples. In the previous work, the data was measured only in X quadrant and reflected about symmetry planes to simply show the 2D character of Fermi surface that is different from 1D chain. In this work, we measured both X and Y directions on the same cleave to compare these in the same condition. We could get the same result for several samples cut out from a sample batch with high quality.

In order to understand what causes the significant $a-b$ asymmetry, we carefully
compare the band dispersions along $\Gamma$-Y and $\Gamma$-X.
Figure 3 (a-b) and (c-d) show the EDCs and the corresponding symmetrized EDCs measured along  $\Gamma$-Y and $\Gamma$-X, respectively (see panel (h)).
We determined the energy states of BB (AB) along the two momentum cuts from the peak positions of the EDCs (symmetrized EDCs), and plot these in panel (e).
We find that the BB along $\Gamma$-X has a convex downward dispersion centered at X. This is in contrast with all other dispersions (AB along $\Gamma$-X and both AB and BB along $\Gamma$-Y), which show a convex upward shape. We stress that the former band shape is specific to Y124 because
the latter has been observed in many cuprates \cite{Damascelli} such as Bi$_2$Sr$_2$CaCu$_2$O$_{8+\delta}$ (Bi2212), Bi$_2$Sr$_2$CuO$_{6+\delta}$ (Bi2201), and La$_{2-x}$Sr$_x$CuO$_4$ (LSCO).
Figure 3(g) shows the splitting energy ($2t_ \bot $, where $t_ \bot $ is the interlayer hopping integral within the bi-layers) estimated along $\Gamma$-X and $\Gamma$-Y. It is clear that
$2t_ \bot$ along $\Gamma$-X has a bump close to X and the difference from that along $\Gamma$-X is about $40\%$ ($\sim 60$meV). We therefore conclude that the characteristic band shape close to X (the electron-like Fermi surface in AB and a convex downward dispersion in BB)
is a result of an enhanced bilayer coupling around X, which pushes the AB up above $E_F$ and pushes down the BB more than that around Y.
In Fig. 3(f), we plot the mean-energies of AB and BB along $\Gamma$-X and $\Gamma$-Y.
We find that the dispersion is almost identical along the both cuts,  which indicates that the in-plane orthorhombicity of the Y124 crystal structure ($b/a\approx1.008$) \cite{Karpinski} causes only a negligible  $a$-$b$ axis asymmetry of the $\it{in-plane}$ hopping integral ($t_{//}$). This small crystal orthorhombicity, therefore, is unlikely to cause a significant $a$-$b$ asymmetry of the $\it{interlayer}$ hopping ($t_ \bot $). 
The enhanced bilayer coupling around X, where the Fermi surface of chains is adjacent to that of planes, is most likely due to a non trivial (momentum dependent) coupling between the chains and planes\cite{Yu,Anderson}. It is worth noting that the
current value of $2t_ \bot $ ($\sim$160meV) is about 4 times smaller than that estimated from band calculations ($\sim$600 meV) \cite{Anderson}, which indicates that
strong electron correlations suppress the interlayer hopping.

In conclusion, we find a strong in-plane $a-b$ asymmetry of the electronic structure in YBa$_2$Cu$_4$O$_{8}$ (Y124). The Fermi surface in the antibonding band has a hole-like shape (centered around S) close to Y and an electron-like shape (centered around $\Gamma$) close to X. This asymmetry is caused by an enhanced bilayer band splitting
close to X, where the Fermi surface of chains is adjacent to that of planes, likely due to a nontrivial coupling between the chains and planes. 
These results are crucial to understand the abnormal asymmetries observed in the superconducting gap and the significantly anisotropic spin dynamics, strongly supporting a number of recent theoretical works \cite{Zhou,Eremin,Schnyder}.

We thank O. K. Andersen and J\"org Schmalian for useful remarks. This work was supported by Director Office for Basic Energy Sciences, US DOE and Swiss NCCR MaNEP. Work at the Ames Laboratory was supported by the Department of Energy-Basic Energy Sciences under Contract No. DE-AC02-07CH11358. R. K. gratefully acknowledges support of Swiss National Science Foundation and K. Alex M\"uler Foundation.


\begin{thebibliography}{99}
\bibitem{TB}
We use seven tight-binding fitting functions with $a-b$ axis anisotropy:
$n_0=1$,
$n_1=[(1+\delta_1)\cos k_x+\cos k_y]/2$,
$n_2=\cos k_x\cos k_y$,
$n_3=[(1+\delta_3)\cos 2k_x+\cos 2k_y]/2$,
$n_4=[(1+\delta_4)\cos 2k_x\cos k_y+\cos k_x\cos 2k_y]/2$,
$n_5=\cos 2k_x\cos 2k_y$,
$n_6=[(1+\delta_6)\cos 3k_x+\cos 3k_y]/2$.
The fitting parameter $(t_0 ,t_1 ,t_2 ,t_3 ,t_4 ,t_5 ,t_6 ,\delta _1 ,\delta _3 ,\delta _4 ,\delta _6 )$ for obtained tight-binding band [$\varepsilon (k_x ,k_y ) = \sum {t_i n_i }$] is
(0.242,-0.582,0.189,-0.055,-0.023,-0.007,-0.019,0.029,-0.073,-0.300,0.334) and
(0.100,-0.665,0.166,-0.102,-0.108,-0.009,-0.055,-0.002,-0.072,0.350,-0.056) for antibonding-band and bonding-band, respectively.
\end{thebibliography}
\end{document}